\documentclass{article}
\usepackage{amssymb,amsmath}
\newcommand{\wickn}[1]{\protect{:\!#1\!:}}
\def\1ad{\mbox{\normalsize $^1$}}
\def\2ad{\mbox{\normalsize $^2$}}
\def\3ad{\mbox{\normalsize $^3$}}
\def\4ad{\mbox{\normalsize $^4$}}
\def\5ad{\mbox{\normalsize $^5$}}
\def\6ad{\mbox{\normalsize $^6$}}
\def\7ad{\mbox{\normalsize $^7$}}
\def\8ad{\mbox{\normalsize $^8$}}
\def\makefront{
\vspace*{1cm}\begin{center}
\def\sp{
\renewcommand{\thefootnote}{\fnsymbol{footnote}}
\footnote[4]{corresponding author : \email_speaker}
\renewcommand{\thefootnote}{\arabic{footnote}}
}
\def\newtitleline{\\ \vskip 5pt}
{\Large\bf\titleline}\\
\vskip 1truecm
{\large\bf\authors}\\
\vskip 5truemm
\addresses
\end{center}
\vskip 1truecm
{\bf Abstract:}
\abstracttext
\vskip 1truecm
}
\def\beq{\begin{equation}}                     % 
\def\eeq{\end{equation}}                       %
\def\bea{\begin{eqnarray}}                     %         %
\def\eea{\end{eqnarray}}                       %       % 
\newcommand{\tsf}[2]{\protect{{\textstyle{\frac{ #1}{#2}}}}}
\newcommand{\ts}[1]{\protect{{\textstyle{ #1}}}}

\newcommand{\msh}[1]{\protect{\frac{d\,{\bf #1}}{2\omega_{\bf #1}}}}

\newcommand{\omsh}[1]{\protect{\omega_{\bf #1}}}

\newcommand{\ra}[1]{\protect{{\bf #1}}}

\begin {document}                 
\def\titleline{
%%%%%%%%%%%%%%%%%%%%%%%%%%%%%%%%%%%%%%%%%%%%%%%
%%%%%%%%%%%%%%%%%%%%%%%%%%%%%%%%%%%%%%%%%%%%%%%
The ultraviolet-finite Hamiltonian approach 
\newtitleline 
on the noncommutative
Minkowski space
%%%%%%%%%%%%%%%%%%%%%%%%%%%%%%%%%%%%%%%%%%%%%%%         
}
\def\email_speaker{
{\tt 
%%%%%%%%%%%%%%%%%%%%%%%%%%%%%%%%%%%%%%%%%%%%%%         
bahns@mail.desy.de       
%%%%%%%%%%%%%%%%%%%%%%%%%%%%%%%%%%%%%%%%%%%%%%
}}
\def\authors{
%%%%%%%%%%%%%%%%%%%%%%%%%%%%%%%%%%%%%%%%%%%%%%
Dorothea Bahns\sp
%%%%%%%%%%%%%%%%%%%%%%%%%%%%%%%%%%%%%%%%%%%%%%
}
\def\addresses{
%%%%%%%%%%%%%%%%%%%%%%%%%%%%%%%%%%%%%%%%%%%%%%
II. Institut f\"ur Theoretische Physik\\
Universit\"at Hamburg\\Luruper Chaussee 149\\
D - 22761 Hamburg
%%%%%%%%%%%%%%%%%%%%%%%%%%%%%%%%%%%%%%%%%%%%%
}

\def\abstracttext{Written version of a talk presented at the 
36th International Symposium Ahrenshoop on the Theory of Elementary Particles, 
26-30 August 2003, Wernsdorf, Germany.

This is an exposition of joint work with S.~Doplicher,
K.~Fredenhagen, and G.~Piacitelli on field theory on the noncommutative
Minkowski space~\cite{BDFPdiag}. The limit of coinciding points is modified
compared to ordinary field theory in a suitable way which allows for the
definition of so-called regularized field monomials as interaction terms.
Employing these  in the Hamiltonian formalism results in
an ultraviolet finite $S$-matrix. }

\large
\makefront

%%%%%%%%%%%%%%%%%%%%%%%%%%%%%%%%%%%%%%%%%%%%%%%%
%%%%%%%%%%%%%%%%%%%%%%%%%%%%%%%%%%%%%%%%%%%%%%%%

\section{Introduction}

Noncommutative spacetimes are studied for various reasons, one of them being
the thought experiment that Heisenberg's uncertainty relation in conjunction with the
laws of classical gravity leads to a restriction as to the best possible
localization of an event in spacetime. The idea is that the simultaneous
measurement of two or more spacetime directions with an arbitrarily high
precision requires an arbitrarily high energy which could
result in building a horizon, cf. for instance~\cite{dfr}. Another
motivation is based on string theory, where field theories on
noncommutative spacetimes are derived as special low-energy limits of open
string theories on $D$-brane configurations in background magnetic
fields~\cite{schomerus,seibwitten}.

The model on which our analysis is founded was defined in~\cite{dfr}, where
continuous spacetime is replaced by a noncommutative $C^*$-algebra $\mathcal E$
``generated'' by Hermitean noncommutative coordinate operators $q^0,\ldots,q^3$
with $[q^\mu,q^\nu]=iQ^{\mu\nu}$, $\mu,\nu=0,\dots,3$, subject to so-called
{``quantum conditions''},
\beq\label{cancomm}
Q_{\mu\nu} Q^{\mu\nu}\,=\,0
\,,\qquad
\big({\textstyle\frac{1}{4}}\,Q^{\mu\nu}\,Q^{\rho\sigma}\,
\epsilon_{\,\mu\nu\rho\sigma}\big)^2 
\,=\,\lambda_P^8 \, I
\,,\qquad
[q^\rho,Q^{\mu\nu}]\,=\,0
\eeq
where $\lambda_P$ is the Planck length, $I$ the identity 
(actually, the quantum coordinates,
being unbounded operators, are affiliated to $\mathcal E$, see~\cite{dfr} for
details). The quantum conditions are Poincar\'e invariant, and the commutators are not given by a
fixed matrix. The joint spectrum $\Sigma$ of the
operators~$Q^{\mu\nu}$ is homeomorphic to the non-compact manifold $TS^2 \times
\{1,-1\}$. For any state $\omega$ in the domain of the $[q_\mu,q_\nu]$, the
uncertainties $\Delta_\omega q_\mu = (\omega (q_\mu^2) - \omega
(q_\mu)^2\,)^{-1/2}$ fulfill the following  space-time uncertainty relations:
$
\Delta q_0 \cdot \left (\Delta q_1 +
\Delta q_2 +\Delta q_3\right)  \,\geq\, \lambda_P^2/2
$, $\Delta q_1\cdot\Delta q_2 + \Delta q_1\cdot\Delta q_3  + \Delta q_2 \cdot\Delta
q_3\,\geq\, \lambda_P^2/2
$.
By a generalized Weyl correspondence in the spirit of ordinary quantum
mechanics, the regular realizations of the quantum conditions were found
in~\cite{dfr}, the difficulty being the nontriviality of the spectrum of the
commutators resulting in a nontrivial centre $\mathcal Z$ of (the multiplier
algebra of) $\mathcal E$. In particular, the product of two ``functions of the
quantum coordinates'' is given by the twisted convolution, 
\beq\label{twcon}
f(q) g(q) = \int dk \,dp\,\check f (k) \,\check g(p)\,
e^{-\frac{i}{2}kQp}
\,e^{i(k+p) q}\,,\qquad kQp=k_\mu Q^{\mu\nu}p_\nu\,,\;kq=k_\mu
q^\mu\,,
\eeq
with $f\in \mathcal F L^1(\mathbb R^4)$, $\check f = {\mathcal F}^{-1} f$, 
where $\mathcal F$ is the ordinary Fourier transform. $e^{-\frac{i}{2}kQp}$ is
referred to as the twisting. The full Poincar\'e-group acts as automorphisms on
$\mathcal E$ and derivatives may be defined as the infinitesimal generators of
translations. The evaluation in a point $f(q)\rightarrow f(a)$, $a\in \mathbb
R^4$, fails to be a positive functional. Instead, optimally localized states
$\omega_a$ with localization centre $a\in \mathbb R^4$, minimizing the
uncertainties, have been defined in~\cite{dfr}. Explicitly, for $f$ as above
and $g\in C_0(\Sigma)$, 
\begin{equation}\label{loc_state}
\omega_a (g(Q)f(q))= \int_{\Sigma_1}d\mu_\sigma
g(\sigma)\int dk\;\check f(k)\,\omega_a(e^{ikq})
= \int_{\Sigma_1}d\mu_\sigma
g(\sigma)\int dk\;\check f(k)\,e^{-\frac {1}{2}|k|^2 }\,e^{ika}
\,,
\end{equation}
where $|k|^2={k_0}^2+\dots+{k_3}^2$ and where  $\mu$ is any probability measure
on a distinguished subset $\Sigma_1\subset\Sigma$. The definition is rotation-
and translation-invariant but not invariant under boosts. This is a problem in
general, as states on $\mathcal E$ must take values in $\mathbb C$, and,
therefore, one has to get rid of the dependence on the commutators by
integrating with respect to some measure on $\Sigma$.  Unfortunately, the
Lorentz group is not amenable, and there is no obvious Lorentz-invariant measure
on $\Sigma$. In analogy with the definition of $f(q)$, a free quantum field
$\phi(q)$ on the noncommutative Minkowski space   was defined in~\cite{dfr} as
\beq\label{field}
\phi(q)\stackrel{\rm def}{=}\int dk\;  \check\phi(k)\otimes e^{ikq}
= (2\pi)^{-3/2}\int \msh k 
\;\big(\,a({ k})\,\otimes \,e^{-ikq}+a^*({k})\,\otimes
\,e^{ikq}\,\big)\,\big|_{k\in
H_m^+}
\eeq
with ordinary annihilation and creation operators $a$ and $a^*$, $\omsh k
= \sqrt{\ra k^2 +m^2}$, and the  ordinary positive mass-shell $H_m^+$. The
field $\phi$ is to be interpreted as a linear map from states on $\mathcal E$
to smeared field operators, 
$\omega\mapsto\phi(\omega)=\langle I\otimes \omega,\;\phi(q)\rangle=
\int dx\;\phi(x)\,\psi_\omega(x)$, 
where on the right-hand side, $\phi(x)$ is a quantum field on ordinary
spacetime, smeared with a testfunction $\psi_\omega$ defined by
$\check\psi_{\omega}(k) = \omega ( e^{ikq})$.  

In last year's talk I spoke about our results on the problem of
unitarity~\cite{BDFPuni}, pointing out that a careful definition of the
time-ordering leads to unitary perturbative setups for general noncommutative
spacetimes, making the restriction to lightlike or space-space noncommutativity
unnecessary. Two different unitary setups were discussed, the Hamiltonian
approach already proposed in~\cite{dfr} and the Yang-Feldman approach, and it
was pointed out in particular, that the internal lines in these unitary
approaches,  are not, in general, given by  Feynman propagators (unless
lightlike or space-space noncommutativity are assumed, in which cases the two
unitary approaches  coincide with the modified Feynman rules).

The interaction term employed was a normally ordered product
$\wickn{\phi^n(q)}$ as proposed in~\cite{dfr}.  While this is a straightforward
generalization of the ordinary local interaction term~$\wickn{\phi^n(x)}$, it
is not the only possibility,  and one of the important questions in the field
is how ordinary local interaction terms are to be replaced in the
noncommutative setting. While plagued with problems such as the violation of
causality, field theories on the noncommutative Minkowski space do allow for
some notions of locality from which suitable interaction terms may be derived.
One of the possibilities is a programme adapted to the Yang-Feldman
approach resulting in the introduction of the so-called quasiplanar Wick
products~\cite{BDFPquasi,bahnsdiss}.
Another possibility, which was elaborated
in~\cite{BDFPdiag}, is the topic of this talk. It is more natural in the
Hamiltonian approach, and is based on re-defining the concept of  products of
fields evaluated at the same point. Since, by construction, strict localization
on the noncommutative Minkowski space is impossible, it is argued that fields
cannot be evaluated ``at the same point'', but only at points which are ``close
together''. Using the optimally localized states, this notion is made precise
and the so-called quantum diagonal map is introduced. Applying this map to
define interaction terms, Hamilton operators are found which lead to
ultraviolet finite theories for any $\phi^n$-self-interaction, see~\cite{BDFPdiag}.
An alternative proof is sketched here, see~\cite{bahnsdiss}
for details.

%%%%%%%%%%%%%%%%%%%%%%%%%%%%%%%%

\section{An approximate limit of coinciding points}

Starting point are mutually commuting sets of quantum coordinates, 
$q^\mu_j=I\otimes \dots \otimes I\otimes q^\mu\otimes I\otimes \dots\otimes I$,
$j=1,\dots ,n$ with $q^\mu$ in the $j$-th tensor factor. As in field theory on
the ordinary Minkowski space, the product of fields at such different
``points'', 
\[
\phi(q_1)\dots\phi(q_n)=(2\pi)^{-4n}\int dk_1\dots dk_n\, 
\check \phi(k_1)\dots\check \phi(k_n)\,e^{ik_1q_1}\dots e^{ik_n q_n} \,,
\] 
is well-defined, mapping states to field operators on the ordinary Fock space.
Starting from this expression, one may try to give
meaning to a field monomial in coinciding points in order to define the
generalization of a local interaction term. The exact limit of coinciding
points cannot be assumed, since, contrary to the ordinary case, the relative
coordinates $q_{ij}^\mu=q^\mu_i-q^\mu_j$ do not all commute with each other,
making it impossible to set all differences to zero simultaneously. Instead,
one may define an ``approximate'' limit of coinciding
points~\cite{hesselb,gherdiss} and
minimize the relative coordinates using the states with minimal uncertainty
(optimal localization) with localization centre $a=0$. For a related discussion
see also~\cite{madore}. A simplification used in the
construction is to take the tensor product in the definition of the mutually
commuting sets of coordinates not over the complex numbers $\mathbb C$ but over
the centre $\mathcal Z$, 
\begin{equation}\label{tensZq}
q^\mu_i\stackrel{\rm def}{=}I\otimes_{\mathcal Z} \dots \otimes_{\mathcal Z}
 I\otimes_{\mathcal Z} q^\mu\otimes_{\mathcal Z}I\otimes_{\mathcal Z}
 \dots\otimes_{\mathcal Z} I  \,,
\end{equation}
which means, in particular, that the tensor product is linear with respect to
twistings.  The  quantum coordinates $q_j$ then satisfy the canonical
commutation relations,
\[
[q_j^\mu,q_k^\nu]=0 \mbox{ for } j\neq k\,,\qquad 
[q_j^\mu,q_j^\nu]=i\,Q^{\mu\nu} 
%\otimes_{\mathcal Z}I\otimes_{\mathcal Z} \dots \otimes_{\mathcal Z} I\,,
\]
where the right-hand side does not depend on $j$, such that, in particular,
differences of commutators $[q^\mu_j,q^\nu_j]-[q^\mu_k,q^\nu_k]$  are zero.
Here, the $Q^{\mu\nu}$ are subject to the quantum conditions~(\ref{cancomm}). 
Employing the tensor product over $\mathcal Z$  also implies that the mean
coordinates $\bar q^\mu=\tsf 1 n (q^\mu_1+\dots+q^\mu_n)$  commute with the
relative coordinates, and behave like quantum coordinates of characteristic
length $1/\sqrt{n}$, i.e. $[\bar q^\mu,\bar q^\nu] =i\,\tsf 1 n Q^{\mu\nu}$. 

Each quantum  coordinate $q^\mu_j$ is then rewritten in terms of the mean
coordinate and the relative coordinates, and a so-called quantum
diagonal map $E^{(n)}$ is be defined, which minimizes all relative
coordinates using the states of optimal localization, while leaving the mean
coordinate invariant. For the field monomial we then find
explicitly, cf.~\cite{BDFPdiag}:
\[
\phi^n_R(\mathfrak q)\stackrel{\rm def}{=}E^{(n)}(\phi(q_1)\dotsc\phi( q_n))=
\int dk_1\dots dk_n \;\check \phi(k_1)\dots\check \phi(k_n)\,r_n(k_1,\dots,k_n)
\,e^{i\left(\sum_ik_i\right){\mathfrak q}}\,,
\]
where the quantum coordinates $\mathfrak q^\mu$ with characteristic length
$1/\sqrt{n}$ correspond to the mean coordinates, and where the kernel $r_n$ is
given by 
\begin{equation}
\label{rn}
r_n(k_1,\dotsc,k_n)=\exp\big(-\tsf{1}{2}\;
\ts{\sum\limits_{i=1}^n}\;\big|{k_i}
-\tsf{1}{n} \;\ts{\sum\limits_{l=1}^n}\; {k_l}\big|^2\;\big)\,.
\end{equation}
The Gaussian factors result from the application of the states of optimal
localization to the relative coordinates. Note that no twistings appear due to
the fact that we started from mutually commuting coordinates\footnote{To be
exact, the reader is reminded that the application of  states of optimal
localization also involves an integration with respect to some measure on
$\Sigma_1$. However, $E^{(n)}(\phi(q_1)\dots \phi(q_n))$ is constant as a
function of $\sigma\in\Sigma_1$ and no explicit dependence on $\sigma\in
\Sigma_1$ appears.}.

\section{Dyson's series}

This definition is now applied in an effective perturbation  theory on the
ordinary Minkowski space based on the ordinary Dyson series and an effective
nonlocal Hamilton operator. Starting point is the symbol
$\phi_R^n(x)$ of $\phi_R^n(\mathfrak q)$, $\phi_R^n(\mathfrak
q)=(2\pi)^{-4}\int dk\; e^{ik\mathfrak q}\,\int dx\,e^{-ikx}\, \phi_R^n(x)$, for
which we find
\beq\label{phireg}
\phi_R^n(x)= c_n\int da_1\dots da_n\,
\exp\big(-\tsf{1}{2}\ts{\sum\limits_{j=1}^n}  |a_j-x|^2\big)\,
\delta^{(4)}\big(\ts{\sum\limits_{j=1}^n}a_j-n\,x\,\big)\,\phi(a_1)\dotsm
\phi(a_n)\,.
\eeq
Due to the Gaussian functions, the operator valued distribution $\phi_R^n(x)$ 
may be called a {\em regularized field monomial}, since
the evaluation in a testfunction $g\in \mathcal S(\mathbb R^4)$, 
\[
\int dx \,g(x)\,\phi_R^n(x)=\int  da_1\dots da_n\;
g(\,{\textstyle \frac{1}{n}\sum\limits_{i=1}^n a_i})\,\prod_{j=1}^n
\exp\big({-\tsf{1}{2}|a_j-\textstyle{\frac{1}{n}\sum\limits_{i=1}^n a_i}
|^2}\big) 
\; \phi(a_1)\dots\phi(a_n)\,,
\]
is well-defined. This follows directly from the fact that the product of $g$
and the  Gaussian functions as above provides a testfunction on $\mathbb
R^{4n}$, and that the tensor product of fields is always well-defined. Note
that the Gaussian functions alone do not yield a testfunction on $\mathbb
R^{4n}$ as the arguments are not linearly independent, their sum being
zero. For details see~\cite{bahnsdiss}.
In other words, there is no need to bring the
annihilation and creation operators in the regularized field monomial
$\phi_R^n(x)$ into normal order as in the ordinary case, and all tadpoles turn
out to be finite. Nevertheless, the
effective Hamiltonian is  defined with a normally ordered interaction, 
\begin{equation}\label{ham}
H^{g}_{I}(t)
=\tsf 1 {n!} \,\int dx\,\delta(x_0-t)\,  g(x)\,\wickn{\phi_R^n(x)}\,,
\end{equation}
where the regularized Wick monomial $\wickn{\phi_R^n(x)}$ is defined as
in~(\ref{phireg}), but with a normally ordered tensor product of fields, 
$\wickn{\phi(a_1)\dotsm \phi(a_n)}$. Here, an adiabatic switching $g\in
\mathcal S(\mathbb R^4)$, which does not have  a counterpart on the
noncommutative side, is applied to postpone questions concerning the
infrared-behaviour of the theory. Using the regularized field monomials also in
the definition of the free Hamiltonian density would result in a modified
Hamilton operator which is no longer the zero component of a Lorentz vector.
Hence, the free part of the theory and the interaction term are treated on
different footings, and the former is assumed to be governed by the free
Hamiltonian on the noncommutative Minkowski space  introduced in~\cite{dfr},
which turned out to be the same as the one on the ordinary Minkowski space.  The
free part of the theory requiring normal ordering makes it natural to assume
the same for the interaction term. For
a general discussion of the Hamiltonian approach on noncommutative spacetimes,
including a discussion of the resulting graph theory and its equivalence to the
so-called interaction point time ordering approach, see~\cite{bahnsdiss}.

Following the proposal from~\cite{dfr}, the $S$-matrix corresponding to the
interaction Hamilton operator $H^g_I(t)$ was defined in~\cite{BDFPdiag},
employing the ordinary Dyson series, 
$
S[g]=
I+\sum_{r=1}^\infty S_r[g]
$,
where 
\bea\label{SNregHam}
S_r[g]&=& \left(\tsf{-i}{n!}\right)^r\,
\int dt_1\dots dt_r\,\theta(t_{1}-t_{2})\dots
\theta(t_{{r-1}}-t_{r})\,H^g_I(t_1)\dots H^g_I(t_r)\,.
\eea
Note that $S$ is formally unitary by the fact that  the effective Hamiltonian
is symmetric, $H^{g}_{I}(t)^* = H^{g}_{I}(t)$. Moreover, as we have shown
in~\cite{BDFPdiag}, due to the regularizing Gaussian kernel, the $S$-matrix
$S[g]$ is well-defined in every order~$r$.  No ultraviolet divergences appear.

As an alternative to the analysis in~\cite{BDFPdiag}, a proof in position space
can be given, see~\cite{bahnsdiss}. The idea is the following. First consider
ordinary quantum field theory, where the $S$-matrix at $r$-th order is given by
\[
S_r[g]=c
\int dx^1\dotsm dx^r \;g(x^1)
 \dots g(x^r) \; \prod_{j=1}^{r-1}\,\theta\big(x^j_0-x^{j+1}_0 \big)
\,\prod_{j=1}^r \wickn{\phi(x^j)^n}\,.
\]
Expectation values in multi-particle states (without smearing
in the momenta) are typically of the form $
\prod_{j<j^\prime}
\Delta_+(x^j-x^{j^\prime})^{n(j,j^\prime)}\; \langle p_{(l)} |\,
\wickn{
\phi(x^1)^{m_1}\dots \phi(x^r)^{m_r}}\,|q_{(s)}\rangle
$
for some choice of indices $j,j^\prime\in R=\{1,\dots,r\}$. Here,
$n(j,j^\prime)=n(j^\prime,j)\in\mathbb N_0$ and 
$m_j=n-\sum_{i\in R} n(j,i)$. While the
multiplication of a translation-invariant distribution with a Wick product of
fields is well-defined, ultraviolet
divergences arise since the product of a Heaviside function $\theta$ with
contractions $\Delta_+^n$ is ill-defined in $0$ for $n \geq 2$. 
Such divergences do not appear if regularized Wick monomials are employed. 
Here, we find the following expression for~$S_r[g]$:
\bea\label{Sregpos}
S_r[g]&=&c
\int d\underline{a}^1\dotsm d\underline{a}^r
\;g(\kappa(\underline{a}^1))
 \dots g(\kappa(\underline{a}^r)) \;
\prod_{j=1}^{r-1}\,\theta\big(\kappa_0(\underline{a}^{j})
-\kappa_0(\underline{a}^{j+1})\big)
\nonumber \\&&\qquad\quad\cdot
\,\prod_{j=1}^r \,\big(\;
\exp\big({-\tsf{1}{2}|\underline{a}^j-\kappa(\underline{a}^j)|^2}\big)\;
\wickn{\phi(a^j_1)\dots\phi(a^j_n)}\;\big)\,,
\eea
where $\underline{a}=(a_1,\dots,a_n)$, $\underline{a}-x=
(a_1-x,a_2-x,\cdots,a_n-x)$, $d\underline{a}=da_1\dots da_n$  for
$x,a_j\in\mathbb R^4$, and where  $\kappa(\underline{a})\in\mathbb R^4$ is the
mean of $\underline{a}$, $\kappa(\underline{a})
=\frac{1}{n}(a_1+\dots+a_n)$, and has the time component
$\kappa_0(\underline{a})=\frac{1}{n}(a_{1,0}+\dots+a_{n,0})$.
Taking expectation values of $S_r[g]$ in multi-particle states (without
smearing in the momenta) then yields contractions which appear {\em exactly
once}. Hence, multiplication with the Heaviside functions does not
pose a problem. Moreover, as the product of Heaviside functions 
in~(\ref{Sregpos}) is a translation-invariant distribution, 
its multiplication with the Wick product of fields is not problematic. And
as the arguments of the contractions differ from those appearing in
the fields, the product of contractions and fields is actually a tensor product
and as such automatically well-defined. Furthermore, by a similar argument as employed
in the discussion of the regularized field monomials, one concludes that the
Gaussian functions together with the $r$ adiabatic switching functions at
$r$-th order  yield a  testfunction on $\mathbb R^{4nr}$. This proves the
claim.

If no normal ordering is employed in the definition of the interaction
Hamiltonian, the $S$-matrix is still finite at every order of the perturbative
expansion as long as an adiabatic switching~$g$ is employed. An analysis of the
adiabtic limit where $g$ is replaced by a constant may be found
in~\cite{bahnsdiss}. For instance, the  vacuum expectation values of the
$S$-matrix~(\ref{SNregHam}) diverge in this case. Furthermore, certain tadpoles
turn out to diverge which provides further motivation as to why to consider the
normally ordered interaction term. Moreover, the spatial cutoff may
safely be removed in
non-vacuum graphs, if the
normally ordered interaction term is applied. Trying to remove the time-cutoff,
however, one encounters a peculiar kind of divergence in particular graphs
which are not one-particle irreducible. Such divergences will
be absent when dressed propagators are employed - for which there is no need
from the point of view of ultraviolet divergences. This may be seen as a subtle
form of the ultraviolet-infrared mixing problem which otherwise is not present
in the approach presented here, since the Gaussian kernels never cancel. For more details see~\cite{bahnsdiss}.

\section{Conclusion}

The regularized Wick products can also be applied to theories with constant
noncommutativity matrix $\theta$, yielding  an ultraviolet finite $S$-matrix
for $\theta$ with maximal rank. However, if $\theta$ is degenerate  as is the
case in theories with commuting time, the best-localized states may not suffice
as a regularization. Major drawbacks of the approach  presented here (apart from
fundamental problems such as the well-posedness of the initial value problem)
are that the definition of the best localized states and hence that of the
quantum diagonal map breaks Lorentz covariance, although rotation and
translation invariance are kept, and that the free part of the theory is
treated on  a different footing than the interaction term. Nonetheless, the
regularized Wick monomials do
provide a framework in which one of the original aims of noncommutative
spacetimes, namely to yield ultraviolet finite field theories, can be achieved.

%%%%%%%%%%%%%%%%%%%%%%%%%%%%%%%%%%%%%%%%%%%%%%%%

\end{document}